\documentclass{sf2a-conf2016}
\usepackage{graphicx}
\usepackage{hyperref}
\usepackage[]{natbib}  
\usepackage{epstopdf}

\def\BibTeX{{\rm B\kern-.05em{\sc i\kern-.025em b}\kern-.08em
    T\kern-.1667em\lower.7ex\hbox{E}\kern-.125emX}}
\bibpunct{(}{)}{;}{a}{}{,}  

\begin{document}

\TitreGlobal{SF2A 2016}


\title{Follow-up and characterization of the \textit{TESS} exoplanets with SOPHIE, SPIRou, and \textit{JWST} }

\runningtitle{Characterization of the \textit{TESS} exoplanets}

\author{N.~Crouzet}\address{Dunlap Institute for Astronomy and Astrophysics, University of Toronto, 50 St Georges Street, Toronto, ON M4Y 2K2, Canada}

\author{X.~Bonfils}\address{Universit\'e Grenoble Alpes, CNRS, IPAG, 38000 Grenoble, France}
\author{X.~Delfosse$^2$}
\author{I.~Boisse}\address{Aix-Marseille Universit\'e, CNRS, LAM (Laboratoire d'Astrophysique de Marseille) UMR 7326, 13388 Marseille, France}
\author{G.~H\'ebrard}\address{Institut d'Astrophysique de Paris, UMR 7095 CNRS, Universit\'e Pierre \& Marie Curie, 98bis boulevard Arago, 75014 Paris, France}
\author{T.~Forveille$^2$}
\author{J.-F.~Donati}\address{Universit\'e de Toulouse, UPS-OMP, IRAP, 14 avenue E.~Belin, Toulouse, F-31400 France \& CNRS, IRAP/UMR 5277, Toulouse, 14 avenue E.~Belin, F-31400 France}
\author{F.~Bouchy$^{3,}$}\address{Observatoire de Gen\`eve, Universit\'e de Gen\`eve, 51 ch. des Maillettes, 1290 Sauverny, Switzerland}
\author{C.~Moutou$^{3,}$}\address{Canada-France-Hawaii Telescope Corporation, CNRS, 65-1238 Mamalahoa Hwy, Kamuela, HI 96743, USA}
\author{R.~Doyon$^{8,}$}\address{D\'epartement de physique and Observatoire du Mont-M\'egantic, Universit\'e de Montr\'eal, Montr\'eal, QC H3C 3J7, Canada}\address{Institut de Recherche sur les Exoplan\`etes (iREx), Universit\'e de Montr\'eal, D\'epartement de Physique, C.P. 6128 Succ. Centre-ville, Montr\'eal, QC H3C 3J7, Canada}
\author{E.~Artigau$^{8,\,9}$}
\author{L.~Albert$^{8,\,9}$}
\author{L.~Malo$^{7,\,8,\,9}$}

\author{A.~Lecavelier~des~Etangs$^4$}
\author{A.~Santerne$^3$}

\setcounter{page}{237}


\maketitle


\begin{abstract}

The NASA \textit{TESS} mission will deliver hundreds of transiting exoplanet candidates orbiting bright stars. The spectrometers SOPHIE at OHP and SPIRou at CFHT will be ideal to obtain radial velocities of these candidates, confirm their nature, and derive the planets' masses. These measurements will be crucial to deliver the best targets for atmospheric characterization with \textit{JWST}. Here, we calculate the required observing time with SOPHIE, SPIRou, and \textit{JWST} for each of the \textit{TESS} targets in order to prepare follow-up observations. To infer their potential for \textit{JWST}, we restrict the calculations to the case of transmission spectroscopy with NIRISS. The radial velocity follow-up of the giant planets ($R_p > 4 \rm \, R_E$) could be achieved with SOPHIE, with a median observing time of 3.47~hours per target, and a total observing time of 305~hours that includes the 80\% most favorable cases. Several small planets ($R_p < 4 \rm \, R_E$) could also be confirmed, but most of them would require an unrealistic time investment. On the other hand, SPIRou is ideally suited to the follow-up of the small planets, with a median observing time of 2.65 hours per target, and a median observing time of 4.70 hours for the terrestrial planets in the habitable zone ($R_p < 2 \rm \, R_E$, $S < 2 \rm \, S_E$). With \textit{JWST}, the 10\% most favorable small planets (184 planets) have a median observing time of 16.2 hours, and the 10\% most favorable habitable zone terrestrial planets (7 planets) have a median observing time of 59.7 hours. Overall, this study will help define a follow-up strategy and prepare observation programs with SOPHIE and SPIRou before the first planet candidates are delivered by \textit{TESS}.

\end{abstract}



\section{Introduction}

Exoplanets transiting in front of bright stars offer the best prospects for their physical characterization: by delivering more photons, bright stars increase the efficiency of follow-up observations and more information can be obtained on the systems compared to fainter stars. \textit{TESS} \citep[\textit{Transiting Exoplanet Survey Satellite}, launch: 2017, PI:~G.~Ricker,][]{Ricker2014} is a NASA mission to detect transiting exoplanets around bright stars \mbox{($4 < I_{mag} < 13$)} by high precision wide field photometry. One year later, \textit{JWST} (\textit{James Webb Space Telescope}, launch: 2018) will provide a unique opportunity to probe the atmospheres of transiting exoplanets by spectroscopy at infrared wavelengths. Most of the exoplanet targets for \textit{JWST} are expected to come from \textit{TESS}. With such a tight schedule, efficient follow-up of the \textit{TESS} planet candidates is crucial. The spectrometers SOPHIE \citep{Bouchy2009} at OHP (Observatoire de Haute-Provence) and SPIRou \citep{Delfosse2013} at CFHT (Canada-France-Hawaii Telescope) will be ideal to confirm the planetary nature of these candidates and to measure their masses by radial velocities. 
SOPHIE has been successfully used for the follow-up of \textit{CoRoT} and \textit{Kepler} candidates, and the \textit{TESS} targets will be significantly brighter. SOPHIE operates at visible wavelengths (3872-6943 \AA).
SPIRou is a near-infrared spectro-polarimeter that will enable radial velocity measurements at high precision ($<$ 1 $\rm m\,s^{-1}$) and high resolving power ($>$ 70,000) in the 0.98-2.35 $\mu$m bandpass in one shot. One of its main goals is the detection of low-mass exoplanets around low-mass stars, and the first light is planned for the end of 2017, early 2018. The small size of M dwarfs also favours the detection and characterization of small planets through their transits, and  \textit{TESS} will detect many more planets around M dwarfs than \textit{Kepler} and  \textit{K2} thanks to its almost full sky coverage.
In this work, we present estimates of the observing time required with SOPHIE and SPIRou to follow-up the \textit{TESS} simulated planets in radial velocity, and the time required with \textit{JWST} to characterize their atmospheres. This study will be useful to define follow-up strategies and to prepare observation campaigns before the first \textit{TESS} candidates are delivered.
The observation chain and the simulations are described in Section \ref{sec: observation chain}. Section \ref{sec: observing time per target} gives estimates of the required observing time per target. Section \ref{sec: Preparing radial velocity follow-up programs} presents the cumulative observing time with SOPHIE and SPIRou for each planet category in order to prepare follow-up programs.

\section{Observation chain and description of the simulations}
\label{sec: observation chain}

\subsection{Transiting planet candidates with \textit{TESS}}

The main goal of the NASA \textit{TESS} mission is to detect transiting exoplanets orbiting around bright stars. We use the catalog of simulated \textit{TESS} detections presented in \citet{Sullivan2015} as an input of our simulations. We consider the expected yield of planets around the $2\times10^{5}$ target stars: about 70 Earths, 486 Super-Earths, 1111 Sub-Neptunes, and 67 giant planets; these numbers are only indicative as they differ for each simulation. We do not consider detections from the full frame images \citep[see Figure 18 of][for more details]{Sullivan2015}. This catalog provides several physical parameters of the simulated systems: planetary radius $R_p$, orbital period $P$, insolation $S$, host star radius $R_\star$, effective temperature $T_{eff}$, magnitude in several bandpasses, and radial velocity semi-amplitude $K$. We derive other quantities: the planet mass $M_p$ using the empirical mass-radius relation provided by \citet{Weiss2013}, the host star mass $M_s$, the semi-major axis $a$, the transit depth $\delta$ and duration $\tau$, the planet equilibrium temperature $T_{eq}$, density $\rho_p$, surface gravity log$g_p$, and atmospheric scale height $H$. We assume equatorial transits (inclination $i=0$) and circular orbits (eccentricity $e=0$). We do not consider planet multiplicity in this study.

\subsection{Radial velocity follow-up with SOPHIE and SPIRou}

Radial velocity follow-up of the \textit{TESS} candidates will be necessary to confirm their planetary nature and measure their masses. The spectrometers SOPHIE in the visible and SPIRou in the near-infrared are ideal in that purpose. For each \textit{TESS} target, we calculate the observing time $t_{RV}$ that would be necessary to measure $K$ with a given signal to noise ratio $S/N$ with these two instruments. We use a reference radial velocity precision $K_{ref}$ of 6 $\rm m\,s^{-1}$ at a magnitude $V_{ref}$ of 12 in an exposure time $t_{ref}$ of 15 minutes for SOPHIE in the high resolution mode (G.~H\'ebrard, private communication), and an expected precision $K_{ref}$ of 1 $\rm m\,s^{-1}$ at a magnitude $J_{ref}$ of 9.5 in an exposure time $t_{ref}$ of 15 minutes for SPIRou \citep{Delfosse2013}. We scale these reference precisions to the magnitudes of the \textit{TESS} targets according to Poisson noise. We impose a minimum exposure time of 10 minutes to average stellar pulsations, and a floor precision per point of 2 $\rm m\,s^{-1}$ for SOPHIE and 1 $\rm m\,s^{-1}$ for SPIRou. We assume that the measurements are made at quadratures of the radial velocity curves, and we consider that a planet is detected if $K$ is measured with a $S/N$ of 3. In this study, we do not consider stellar rotation and stellar activity which may limit the radial velocity precision.

\subsection{Atmospheric characterization with \textit{JWST}}

The atmosphere of transiting exoplanets can be characterized by transmission and emission spectroscopy during their transits and eclipses, respectively. Here, we consider the case of transit spectroscopy of the \textit{TESS} planets with \textit{JWST}, and we calculate the observing time $t_{JWST}$ that is necessary to detect molecules in their atmospheres. 
We use the \textit{JWST}/NIRISS instrument in the SOSS mode (Single-Object Slitless Spectroscopy) which is dedicated to exoplanet spectroscopy, through the online NIRISS SOSS 1-D Simulator\footnote{\url{http://maestria.astro.umontreal.ca/niriss/simu1D/simu1D.php}}. We run a reference simulation using a set of parameters corresponding to the mean of the \textit{TESS} small planets ($R_p < 4 \rm \, R_E$, where $\rm \, R_E$ is the Earth's radius). Then, we scale the results to each \textit{TESS} target: we calculate the target's flux in the NIRISS bandpass from its $T_{eff}$ and $J$-band magnitude using a blackbody spectral energy distribution, and we assume that the noise scales as the Poisson noise. We assume the same observing time in and out of transit, and we use the 1st order spectrum only. The observation efficiency, defined as integration time / total observing time, is 33\% in the reference simulation. We keep this value for all the targets for simplicity, because it is optimized during the simulation and the simulator can only be ran on individual targets. A future improvement, if available, would be to run the simulator on all the \textit{TESS} targets.
For the atmospheric signal, we calculate the fractional loss of light $\delta_H$ due to molecular absorption from an atmospheric annulus of thickness $H$:
\begin{equation}
\delta_H = \frac{\textrm{d}\delta}{\textrm{d} R_p}\times H = \frac{2HR_p}{R_\star^2}
\end{equation}
This metric allows us to estimate the characteristic amplitude of a molecular absorption feature without relying on a theoretical transmission spectrum. Instead, we consider a single spectral feature with an amplitude $\delta_H$, a width $\Delta\lambda=100$ nm, and a central wavelength $\lambda_c=1.8$ $\mu$m corresponding to the center of the NIRISS SOSS order 1 spectrum. 
Then, we calculate the observing time that is necessary to detect this 1-scale height amplitude feature with a $S/N=1$. In practice, a molecular feature is expected to span several scale heights and a spectrum may contain several features, which would increase the $S/N$ (for example, an amplitude of 3 scale heights would yield $S/N=3$).
These calculations are made for each wavelength tabulated in the NIRISS simulation output and we take the median of the observing times as an estimate for $t_{JWST}$. We also set a lower limit for $t_{JWST}$ as twice the transit duration, because the observations will span at least one full transit including an out-of-transit baseline.

\section{Observing time per target}
\label{sec: observing time per target}

\subsection{SOPHIE and SPIRou}

\begin{figure}[ht!]
 \centering
 \includegraphics[width=8cm,clip]{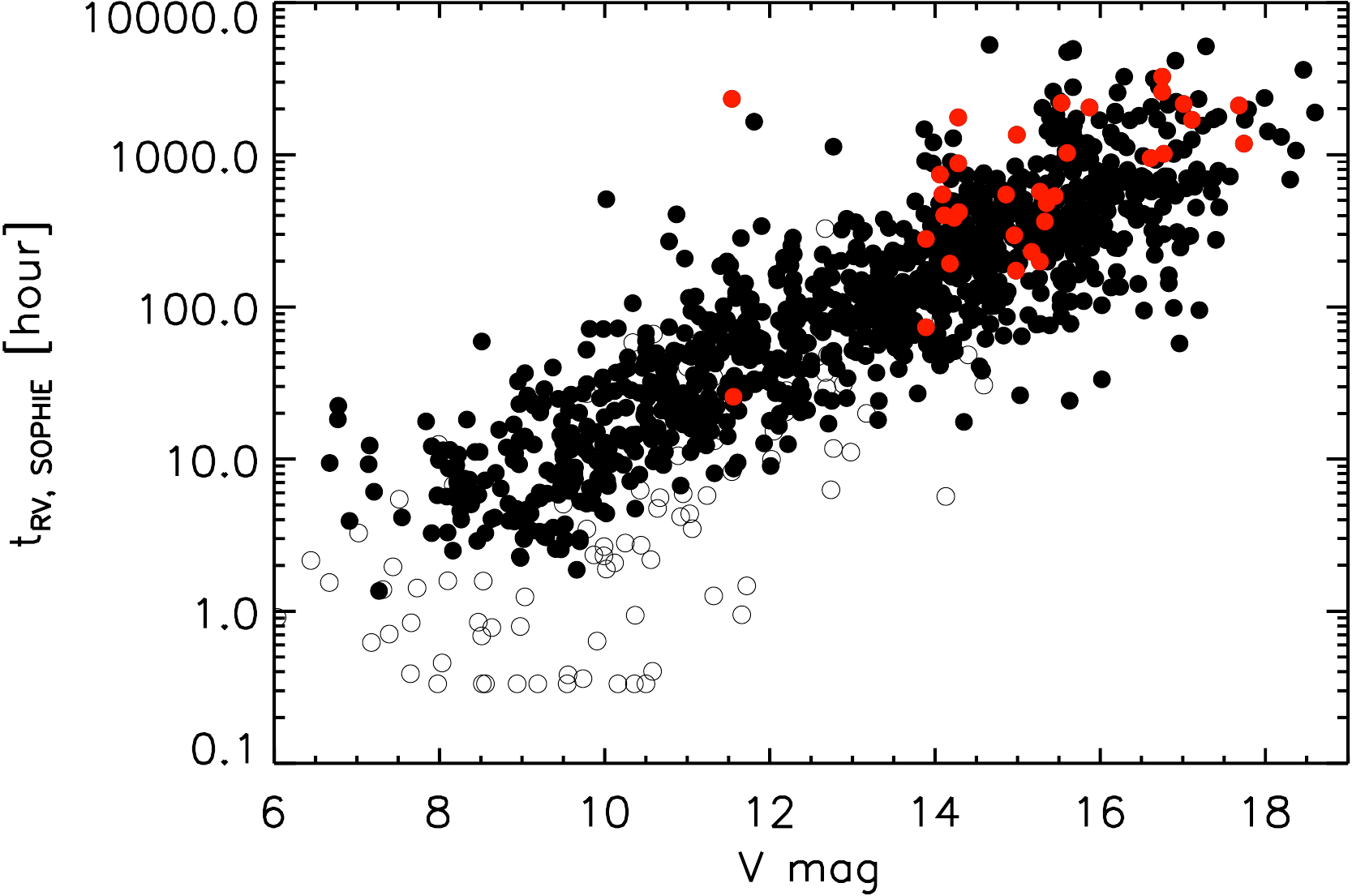}%
 \hspace{0.5cm} \vspace{-0.3cm}
 \includegraphics[width=8cm,clip]{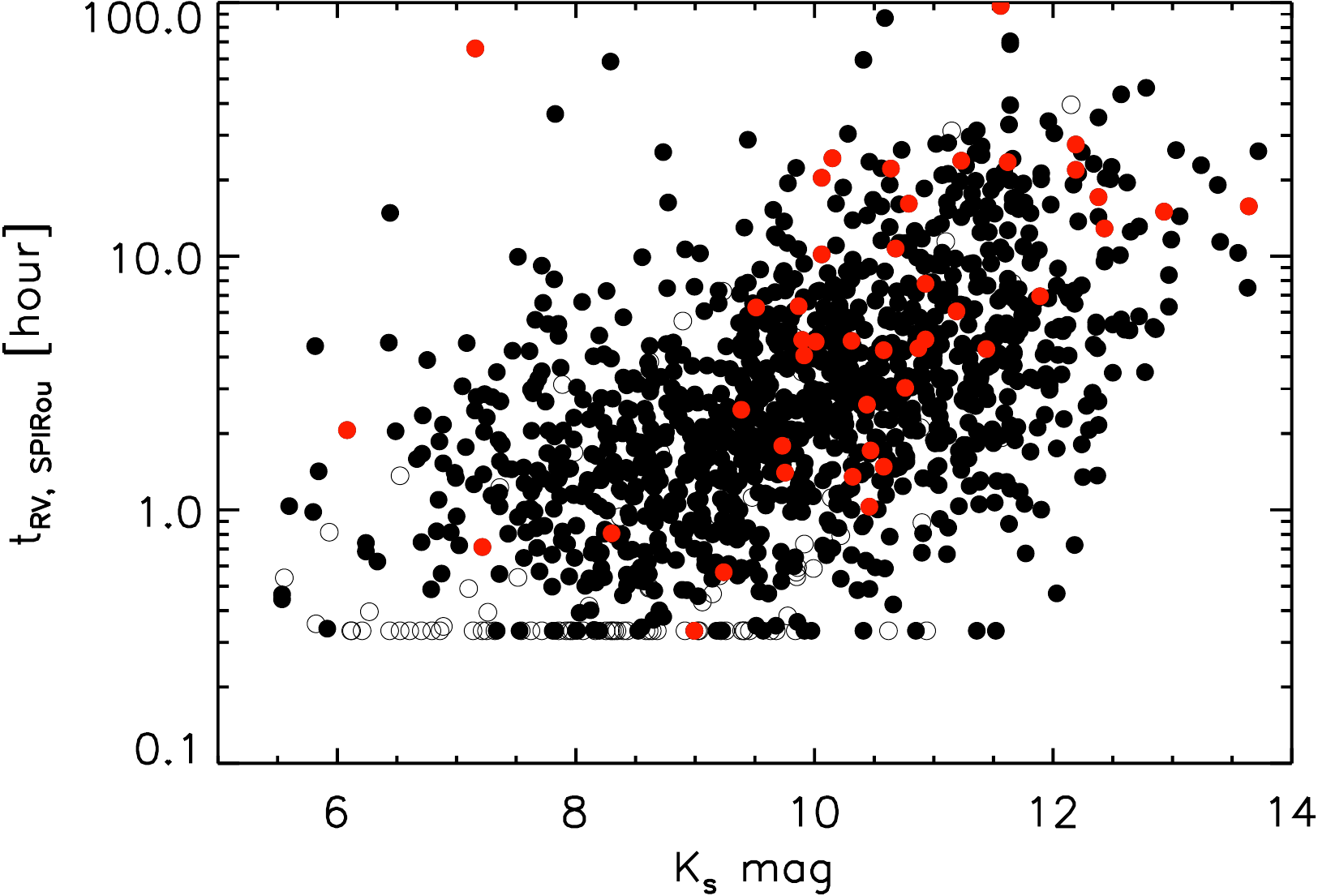}%
  \caption{Observing time required with SOPHIE (left) and SPIRou (right) for each \textit{TESS} target as a function of the $V$ and $K_s$ magnitude, respectively. Giant planets are indicated as open circles, small planets are indicated as filled black circles, and terrestrial planets in the habitable zone are indicated as filled red circles. See text for definitions.}
  \label{fig: time rv}
\end{figure}

We split the target star population into those harbouring giant planets ($R_p > 4 \rm \, R_E$) and those harbouring small planets ($R_p < 4 \rm \, R_E$) including the terrestrial planets in the habitable zone ($R_p < 2 \rm \, R_E$, $S < 2 \rm \, S_E$), where $\rm R_E$ and $\rm S_E$ are the Earth radius and insolation respectively. In this definition, the habitable zone has no outward limit, but most of the \textit{TESS} targets will be observed for 30 days only which will reduce the number of long period, cold planets. We consider only the targets that can be observed with an airmass lower than~2 ($-16.07^\circ < DEC < +90^\circ$ for SOPHIE, $-40.17^\circ < DEC < +79.83^\circ$ for SPIRou).
Figure \ref{fig: time rv} shows the required observing time $t_{RV}$ as a function of stellar magnitude. The follow-up of the giant planets could be achieved with SOPHIE, with a median $t_{RV}$ of 3.47 h per target, and a total time of 305 h after excluding 17 planets with $t_{RV} > 25$~h (the total time varies significantly depending on the limit adopted for $t_{RV}$). Some small planets could also be confirmed by SOPHIE, with a median of 5.92 h and a maximum of 10.71 h for the 10\% most favorable cases (103 planets). However, most of the small planets would require an unrealistic time investment, with a median of 116 h per target. On the other hand, SPIRou is ideally suited to the follow-up of the small planets, with a median $t_{RV}$ of 2.65 h, and a median of only 4.70 h for the terrestrial planets in the habitable zone. As limitations, these calculations do not take into account the observational overheads, and constraining the eccentricity will require taking some data points out of quadratures. Thus, the actual observing times will be slightly larger than the estimates given here.

\subsection{\textit{JWST}}
\label{sec: JWST}

\begin{figure}[ht!]
 \centering
 \includegraphics[width=9cm,clip]{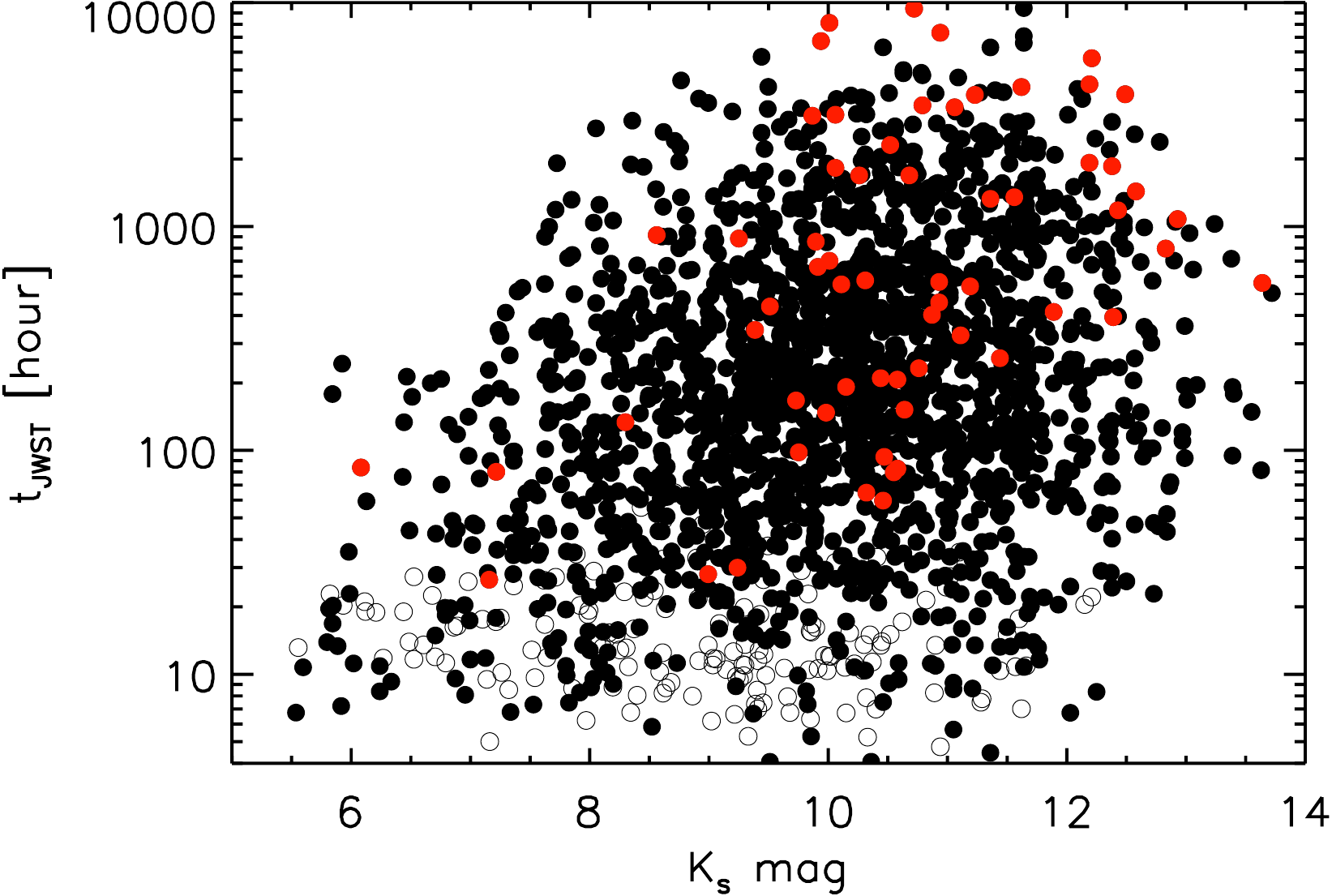}      
 \vspace{-0.3cm}
  \caption{Observing time required with \textit{JWST} NIRISS for each \textit{TESS} planet as a function of the $K_s$ magnitude. The color code is the same as in Figure \ref{fig: time rv}. We set a lower limit on $t_{JWST}$ as twice the transit duration, which is the limiting factor for the giant planets. We show the full \textit{TESS} sample, although targets requiring more than $\sim$100 hours will probably not be observed by \textit{JWST}.}
  \label{fig: time JWST}
\end{figure}

To calculate $H$ and $\delta_H$, we consider a H/He-dominated atmosphere for the giant planets ($\mu=2.32$ u, where $\mu$ is the atmospheric mean molecular mass), a $\rm H_2O$-dominated atmosphere for the hot and warm small planets ($\mu=18$ u), and an Earth-like atmosphere for the cold small planets ($\mu=29$ u). We define hot, warm, and cold according to the planets' insolation: $S > 10 \, \rm S_E$, $2 \, \mathrm{S_E} < S < 10 \, \rm S_E$, and $S < 2 \, \rm S_E$ respectively (with this definition, ``cold'' is equivalent to ``habitable zone''). We consider all the targets, with no restriction on $DEC$.
Figure \ref{fig: time JWST} shows the observing time $t_{JWST}$ required to detect the molecular absorption feature. Interestingly, we find that $t_{JWST}$ increases only weakly with the $K_s$ magnitude for the small planets. This is because $R_\star$ decreases as a function of $K_s$ (M dwarfs are fainter than more massive, larger stars) and as a result $\delta_H$ increases with $K_s$ (the relative transit depth is larger for small stars). We note that $R_\star$ and $R_p$ appear uncorrelated for the \textit{TESS} simulated small planets. Whether this only weak correlation between the time required for transit spectroscopy and the stellar magnitude is generally true for planets around M-dwarfs remains to be investigated.
The median observing time $t_{JWST}$ is 13.5 h for the giant planets and is set by twice the transit duration. The small planets have a median of 201 h, and the 10\% most favorable ones (184 planets) have a median of 16.2 h and a maximum of 28.6 h. The 10\% most favorable habitable zone terrestrial planets (7 planets) have a median of 59.7 h and a maximum of 80.3 h. These planets are highly valuable but will require a significant \textit{JWST} time investment, thus identifying the best ones with SPIRou is crucial. If we assume H/He-dominated atmospheres for all the planets, the median observing time drops to 8 h for the small planets and is limited by twice the transit duration, and to 13.1 h for the habitable zone terrestrial planets.

\section{Preparing radial velocity follow-up programs}
\label{sec: Preparing radial velocity follow-up programs}

\begin{figure}[ht!]
 \centering
 \includegraphics[width=5.5cm,clip]{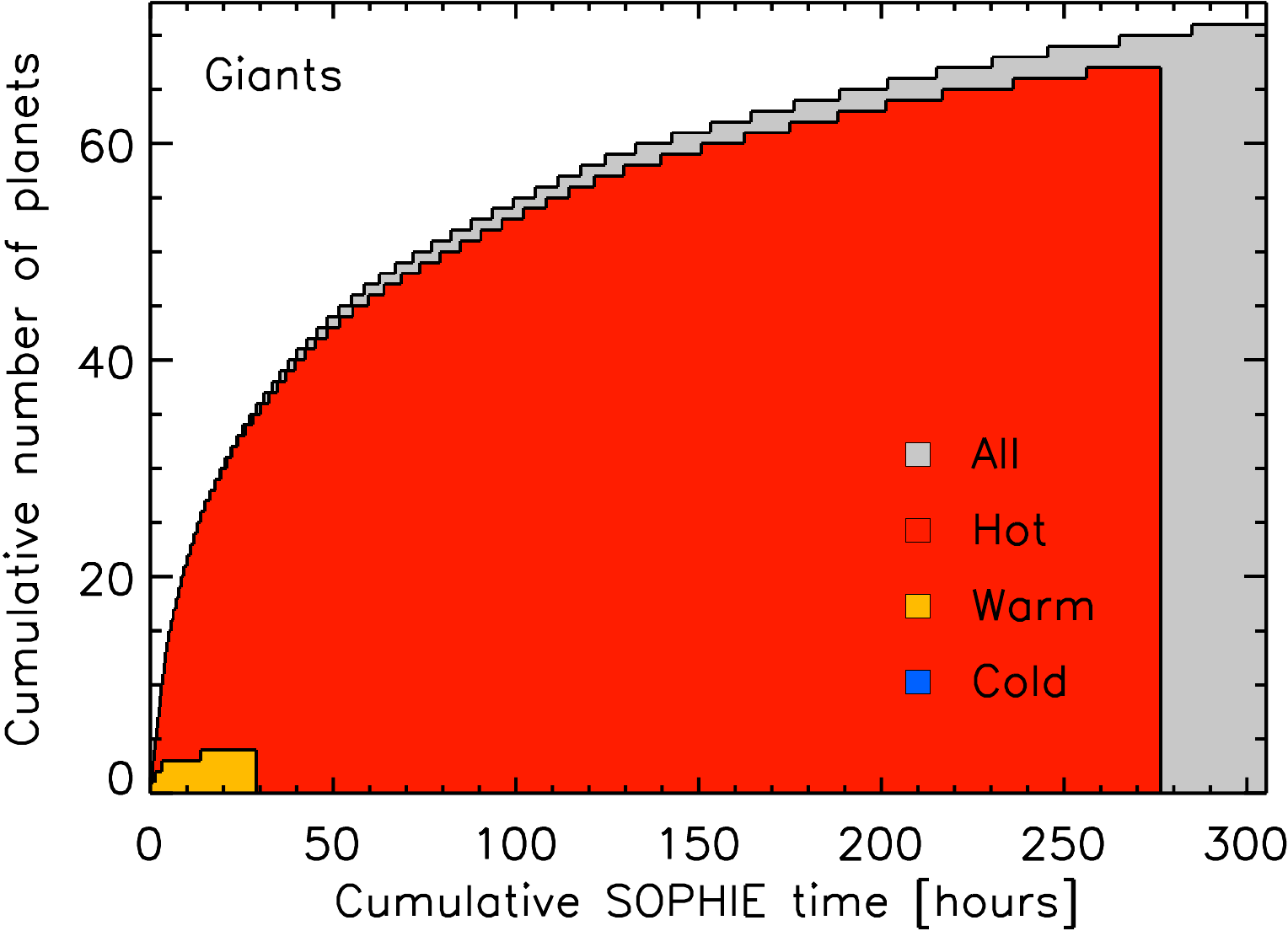}      
 \includegraphics[width=5.5cm,clip]{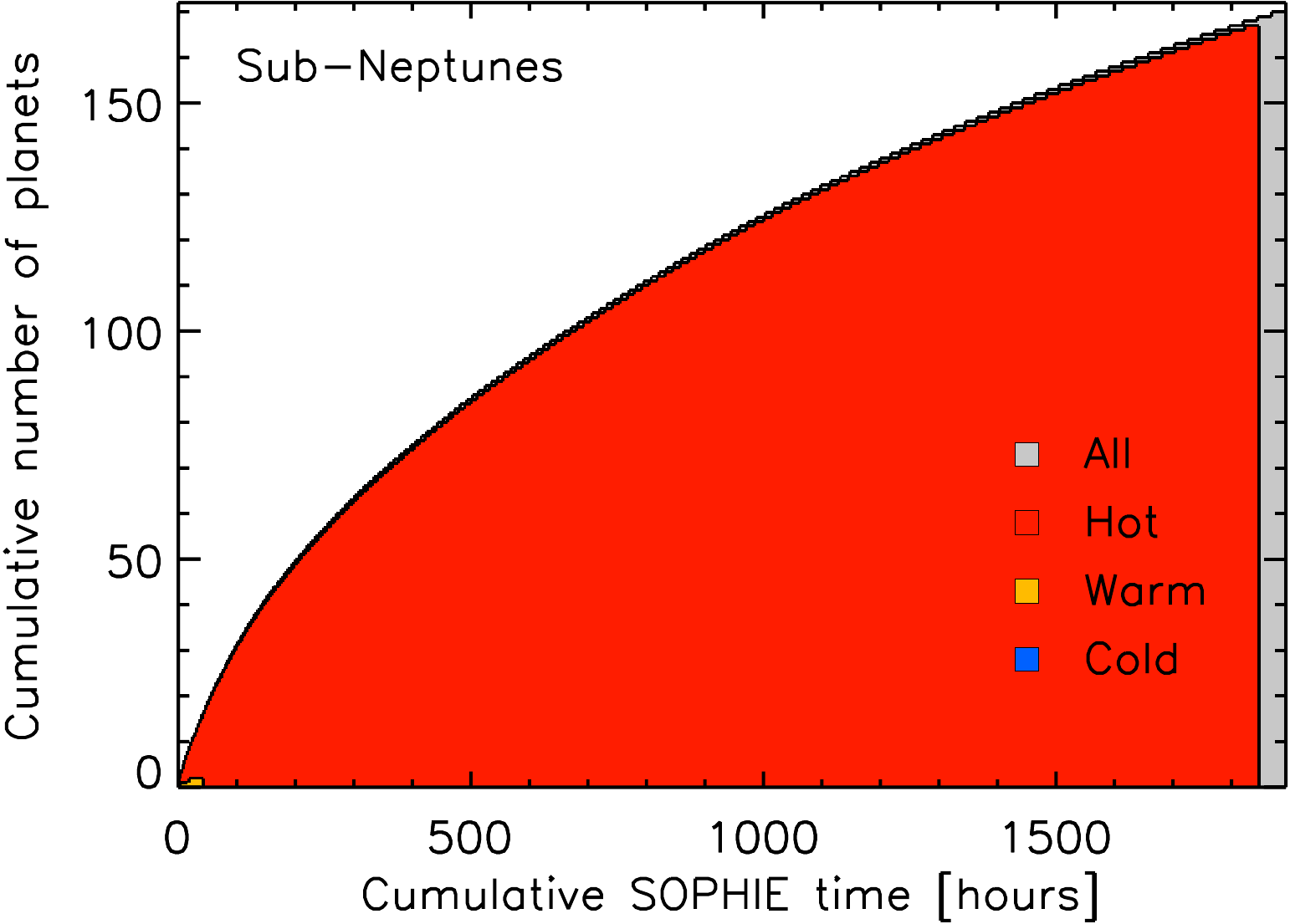}      
 \includegraphics[width=5.4cm,clip]{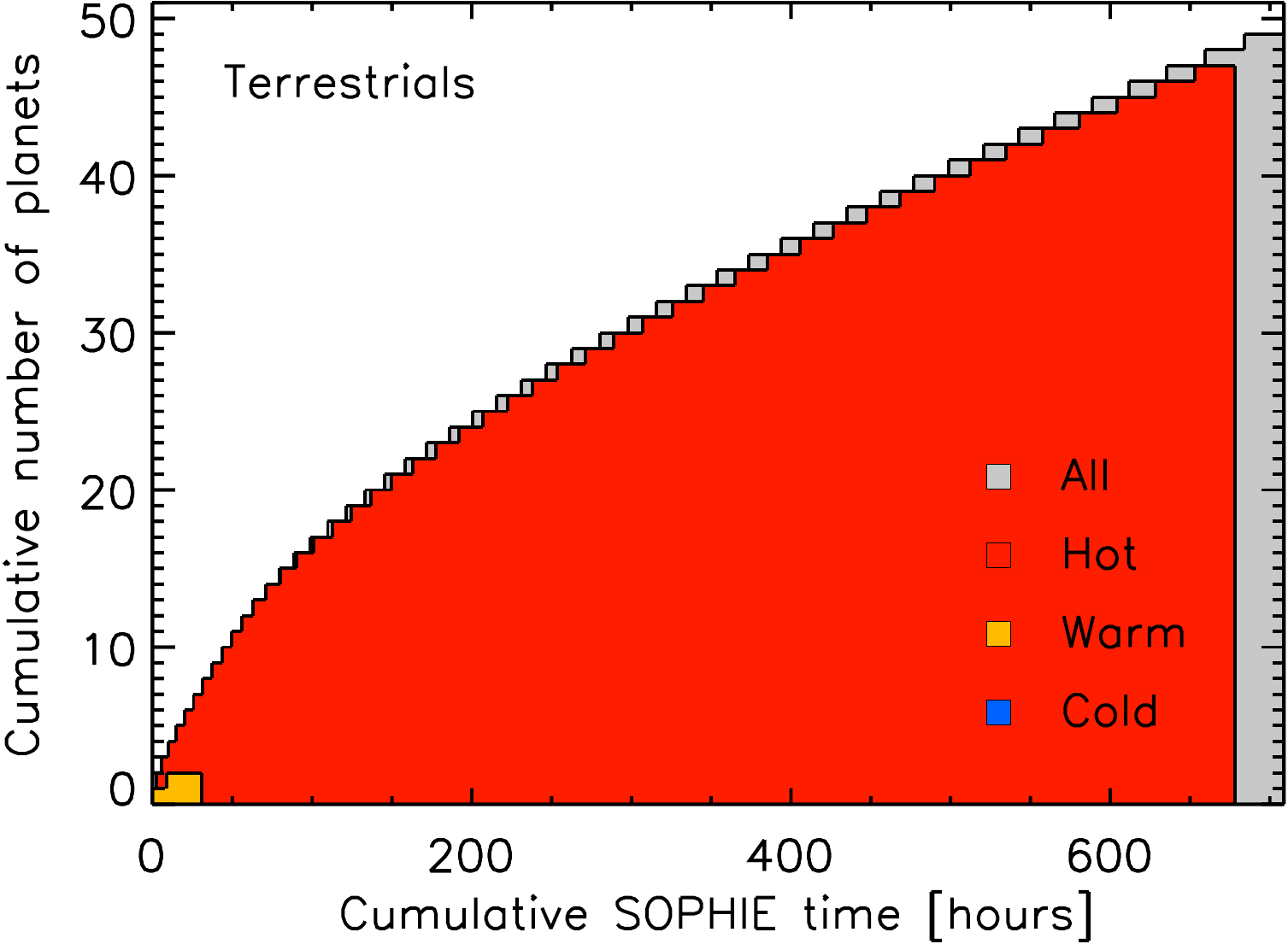}
 \vspace{2.1mm}
 
 \includegraphics[width=5.5cm,clip]{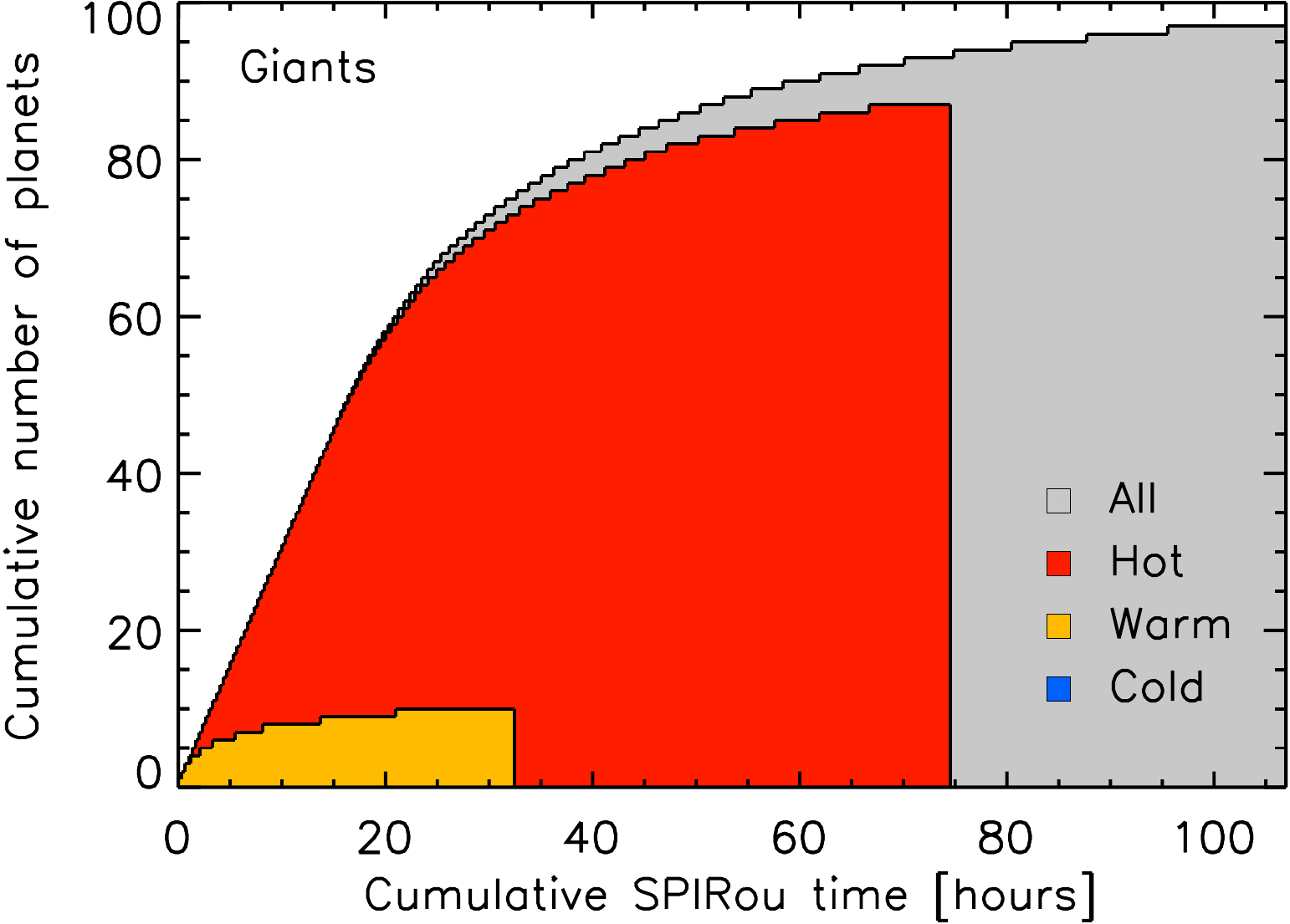}      
 \includegraphics[width=5.5cm,clip]{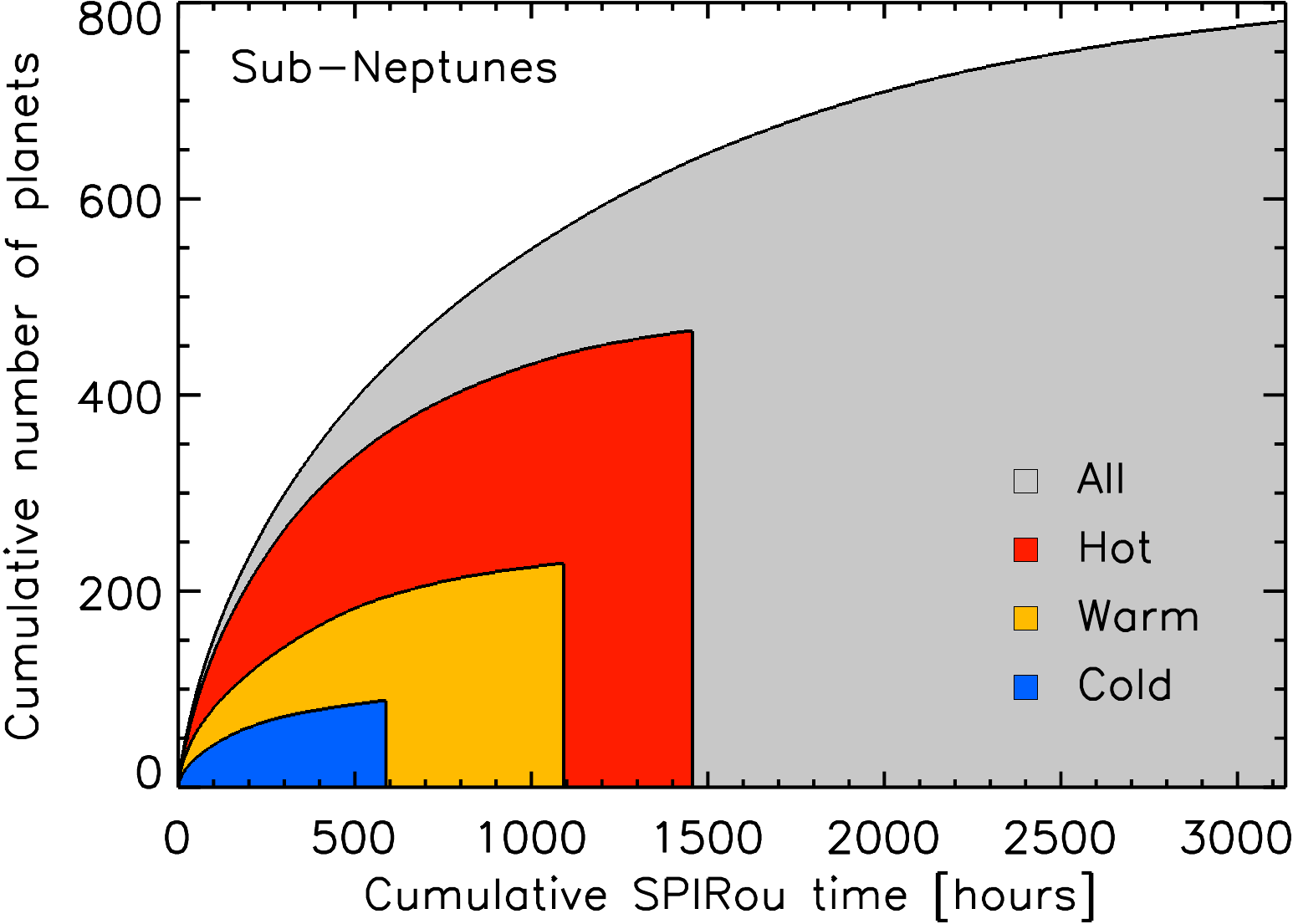}      
 \includegraphics[width=5.5cm,clip]{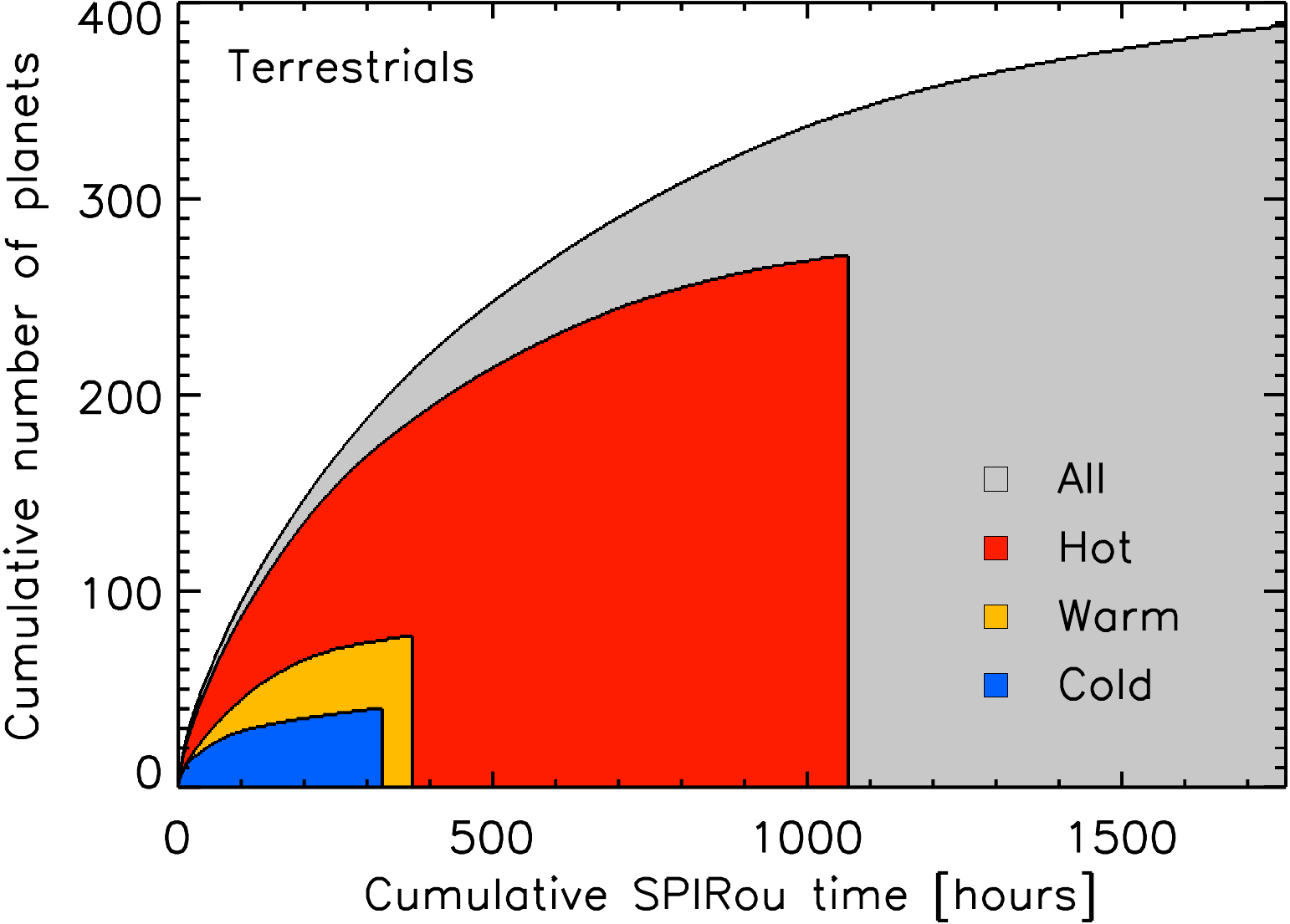} 
 \vspace{-2mm}
 \caption{Cumulative number of \textit{TESS} planets that can be followed-up in radial velocity as a function of the cumulative observing time for SOPHIE (top) and SPIRou (bottom), for the giant planets (left), Sub-Neptunes and Super-Earths (middle), and terrestrial planets (right). We show the full sample (grey), as well as the hot (red), warm (yellow), and cold (blue) samples. The planets are sorted by increasing $t_{RV}$. See text for definitions.}
  \label{fig: cumulative trv}
\end{figure}

In this section, we estimate the cumulative observing time that would be necessary with SOPHIE and SPIRou for the radial velocity follow-up of the \textit{TESS} planet candidates, in order to prepare and optimize observation campaigns. We split the planets into giants ($R_p > 4 \rm \, R_E$), sub-Neptunes and Super-Earths ($2 \, \mathrm{R_E} < R_p < 4 \rm \, R_E$), and terrestrial planets ($R_p < 2 \rm \, R_E$), and into hot, warm, and cold planets as defined in Section \ref{sec: JWST}. We sort the planets by increasing $t_{RV}$ and compute its cumulative distribution. The results are shown in Figure \ref{fig: cumulative trv}. We exclude planets with $t_{RV} > 25$ hours, which would extend the cumulative distributions on the right hand side. These diagrams can be used to define a follow-up strategy: they give directly the number and type of planets that could be followed-up in a given amount of telescope time. Placing the \textit{TESS} candidates on these diagrams as they are being discovered (from an initial guess) will also help decide on their follow-up. Finally, they are useful to define a combined strategy between SOPHIE and SPIRou; an obvious split would be to observe the giant planets with SOPHIE and the small planets around M-dwarfs with SPIRou. These estimates are based on the \textit{TESS} planet simulations: the uncertainties in the number of planets could be as large as 50\% \citep{Sullivan2015} and we do not account for false positive discrimination. As a future improvement, the targets' potential for observations with \textit{JWST} could be included quantitatively in defining the radial velocity follow-up samples, for example by means of a unique merit function taking into account $t_{RV}$ and $t_{JWST}$.

 \vspace{2mm}

\section{Conclusion}

The observing time estimates provided in this study will help define the \textit{TESS} planet candidate samples to observe in radial velocity with SOPHIE and SPIRou as well as their potential for atmospheric characterization with \textit{JWST}. Practically, this approach will help us decide which candidates to observe with SOPHIE and SPIRou while they are being identified by \textit{TESS} and deliver the best targets for \textit{JWST} during the process. This could be particularly important in the context of a large SPIRou program at CFHT, the SPIRou Legacy Survey, which would aim at $\sim$500 nights in the first 5 years of operation. This program would be shared into several science topics and could include the follow-up of the \textit{TESS} low-mass exoplanet candidates around M-dwarfs. Finally, in the coming years, SOPHIE may be upgraded with a new detector in order to improve its sensitivity and extend its wavelength coverage. This would result in a gain equivalent to one magnitude for mid- and early-M-dwarfs.

\begin{acknowledgements}
The Dunlap Institute is funded through an endowment established by the David Dunlap family and the University of Toronto. 
\end{acknowledgements}

\bibliographystyle{aa}  

\begin{thebibliography}{5}
\expandafter\ifx\csname natexlab\endcsname\relax\def\natexlab#1{#1}\fi

\bibitem[{{Bouchy} {et~al.}(2009){Bouchy}, {H{\'e}brard}, {Udry}, {Delfosse},
  {Boisse}, {Desort}, {Bonfils}, {Eggenberger}, {Ehrenreich}, {Forveille},
  {Lagrange}, {Le Coroller}, {Lovis}, {Moutou}, {Pepe}, {Perrier}, {Pont},
  {Queloz}, {Santos}, {S{\'e}gransan}, \& {Vidal-Madjar}}]{Bouchy2009}
{Bouchy}, F., {H{\'e}brard}, G., {Udry}, S., {et~al.} 2009, \aap, 505, 853

\bibitem[{{Delfosse} {et~al.}(2013){Delfosse}, {Donati}, {Kouach},
  {H{\'e}brard}, {Doyon}, {Artigau}, {Bouchy}, {Boisse}, {Brun}, {Hennebelle},
  {Widemann}, {Bouvier}, {Bonfils}, {Morin}, {Moutou}, {Pepe}, {Udry}, {do
  Nascimento}, {Alencar}, {Castilho}, {Martioli}, {Wang}, {Figueira}, \&
  {Santos}}]{Delfosse2013}
{Delfosse}, X., {Donati}, J.-F., {Kouach}, D., {et~al.} 2013, in SF2A-2013:
  Proceedings of the Annual meeting of the French Society of Astronomy and
  Astrophysics, ed. L.~{Cambresy}, F.~{Martins}, E.~{Nuss}, \& A.~{Palacios},
  497--508

\bibitem[{{Ricker} {et~al.}(2014){Ricker}, {Winn}, {Vanderspek}, {Latham},
  {Bakos}, {Bean}, {Berta-Thompson}, {Brown}, {Buchhave}, {Butler}, {Butler},
  {Chaplin}, {Charbonneau}, {Christensen-Dalsgaard}, {Clampin}, {Deming},
  {Doty}, {De Lee}, {Dressing}, {Dunham}, {Endl}, {Fressin}, {Ge}, {Henning},
  {Holman}, {Howard}, {Ida}, {Jenkins}, {Jernigan}, {Johnson}, {Kaltenegger},
  {Kawai}, {Kjeldsen}, {Laughlin}, {Levine}, {Lin}, {Lissauer}, {MacQueen},
  {Marcy}, {McCullough}, {Morton}, {Narita}, {Paegert}, {Palle}, {Pepe},
  {Pepper}, {Quirrenbach}, {Rinehart}, {Sasselov}, {Sato}, {Seager},
  {Sozzetti}, {Stassun}, {Sullivan}, {Szentgyorgyi}, {Torres}, {Udry}, \&
  {Villasenor}}]{Ricker2014}
{Ricker}, G.~R., {Winn}, J.~N., {Vanderspek}, R., {et~al.} 2014, in \procspie,
  Vol. 9143, Space Telescopes and Instrumentation 2014: Optical, Infrared, and
  Millimeter Wave, 914320

\bibitem[{{Sullivan} {et~al.}(2015){Sullivan}, {Winn}, {Berta-Thompson},
  {Charbonneau}, {Deming}, {Dressing}, {Latham}, {Levine}, {McCullough},
  {Morton}, {Ricker}, {Vanderspek}, \& {Woods}}]{Sullivan2015}
{Sullivan}, P.~W., {Winn}, J.~N., {Berta-Thompson}, Z.~K., {et~al.} 2015, \apj,
  809, 77

\bibitem[{{Weiss} {et~al.}(2013){Weiss}, {Marcy}, {Rowe}, {Howard}, {Isaacson},
  {Fortney}, {Miller}, {Demory}, {Fischer}, {Adams}, {Dupree}, {Howell},
  {Kolbl}, {Johnson}, {Horch}, {Everett}, {Fabrycky}, \& {Seager}}]{Weiss2013}
{Weiss}, L.~M., {Marcy}, G.~W., {Rowe}, J.~F., {et~al.} 2013, \apj, 768, 14

\end{thebibliography}

\end{document}